\documentclass[twocolumn,preprintnumbers,amsmath,amssymb]{revtex4}

\usepackage[utf8]{inputenc}
\usepackage{hyperref}
\usepackage{graphicx}
\usepackage{amsmath}
\usepackage{bm}
\usepackage{amssymb}
\usepackage{boxedminipage}
\usepackage{color}
\usepackage{floatrow}

\usepackage{color}

\begin{document}

\title{Guide to optical spectroscopy of layered semiconductors}

\author{Shivangi Shree}
\author{Ioannis Paradisanos}
\author{Xavier Marie}
\author{Cedric Robert}
\author{Bernhard Urbaszek}

\affiliation{\small Universit\'e de Toulouse, INSA-CNRS-UPS, LPCNO, 135 Avenue Rangueil, 31077 Toulouse, France}
\begin{abstract}
In this technical review we give an introduction to optical spectroscopy for layered materials as a powerful, non-invasive tool to access details of the electronic band structure and crystal quality. Potential applications in photonics and optoelectronics are based on our understanding of the light-matter interaction on an atomic monolayer scale. Here atomically thin transition metal dichalcogenides, such as MoS$_2$ and WSe$_2$, are model systems for layered semiconductors with a bandgap in the visible region of the optical spectrum. They can be assembled to form heterostructures and combine the unique properties of the constituent monolayers.  We review the working principles of micro-photoluminescence spectroscopy and optical absorption experiments. We discuss the physical origin of the main absorption and emission features in the optical spectra and how they can be tuned. We explain key-aspects of practical set-ups for performing experiments in different conditions such as variable temperatures or in applied magnetic fields and how parameters such as detection spot size and excitation laser wavelength impact the optical spectra. We describe the important influence of the direct sample environment, such as substrates and encapsulation layers, on the emission and absorption mechanisms. A survey of optical techniques that probe the coupling between layers and analyse carrier polarisation dynamics for spin- and valleytronics is provided.
\end{abstract}

\maketitle
\section{Introduction to layered semiconductors}
\label{optical properties}
\vspace{-0.2cm}
The physical properties of atomic monolayers often change dramatically from those of their parent bulk materials. Prime examples are monolayers of graphite (graphene) and MoS$_2$, as their ultimate thinness makes them extremely promising for applications in electronics and optics. At the same time they give access to new degrees of freedom of the electronic system such as the valley index or interactions between quasi-particles such as excitons (Coulomb bound electron-hole pairs). Additional functionalities emerge as these materials are stacked in van der Waals heterostructures \cite{ubrig2020design}. In addition to the currently investigated materials, recently about 1800 materials were predicted to be exfoliable, stable in monolayer form \cite{mounet2018two} and therefore tools for investigating the properties of these emerging layered materials are of prime importance. 
Here we provide a guide to optical emission and absorption spectroscopy for atomically thin layered materials, commonly carried out in optical microscopes for increased spatial resolution. Optical spectroscopy gives access to key information such as the bandgap, exciton binding energy and absorption strength of a material. Combining spatial and polarisation resolution gives access to the spin and valley physics in monolayers and also in heterostructures. In the latter optical, transitions are tunable over a wide wavelength range and electron-hole pairs can experience nanoscale moir\'e confinement potentials for quantum optics experiments and investigating collective effects of electronic excitations \cite{seyler2019signatures, shimazaki2020strongly}. Moreover, optical spectroscopy techniques can be applied to semiconducting, magnetic layered materials such as chromium trihalides for probing their magnetisation \cite{sun2019giant,ubrig2019low,mak2019probing,paradisanos2020prominent}. Optical spectroscopy also reveals magnetic proximity effects and charge transfer as non-magnetic and magnetic layers are placed in direct contact to form heterostructures \cite{zhong2017van,PhysRevLett.124.197401,lyons2020interplay}. For applications in photonics optical spectroscopy reveals how light matter coupling is enhanced when layered materials are placed in optical cavities or on resonators \cite{sortino2019enhanced,paik2019interlayer}. Optical spectroscopy can be used as a non-invasive technique for studying lattice structure, interlayer coupling and stacking, that provides complementary information to direct atomic-resolution imaging from electron microscopy \cite{zhang2017interlayer,sushko2019high,andersen2019moir,shree2019high,holler2020lowfrequency}.  \\
\indent The target of this review is to give an overview on what kind of information on novel layered materials we can access in practical optical spectroscopy, how the optical spectra are impacted by several distinct parameters such as the set-up and equipment used, the experimental conditions (temperature, external fields), the sample structure and very importantly the active layer environment. The electronic excitations in an atomically thin layer are strongly impacted by the substrate and encapsulating layers. This leads to two directions for experiments, (i) access intrinsic properties of the layers placed in a well controlled environment (i.e. in-between two inert buffer layers) or (ii) the layered material acts as a probe as we make use of the interaction of the optical excitations with the direct environment to investigate, for example, the magnetisation of adjacent layers or detecting molecules in the proximity \cite{zhao2019functionalization}. \\ 
\indent Optical properties of layered semiconductors, using the model system of  transition metal dichalcogenide (TMD) semiconductors are introduced in the remainder of Section \ref{optical properties}, together with fabrication methods for typical sample structures. The equipment used for optical spectroscopy set-ups, commercially available systems or components assembled in a laboratory, are discussed in Section \ref{equipment}. Optical spectroscopy techniques used to uncover the main optical transitions in TMD materials and how they are impacted by the sample structure and the surrounding layers are detailed in the remainder of Section \ref{spectroscopy}.  How to access spin and valley polarisation effects using optical methods is introduced in Section \ref{polarization}. Finally, opportunities for Raman scattering and second harmomic generation (SHG) are outlined in Section \ref{orientation}.

\begin{figure*}[t]
\centering
\includegraphics[width=0.95\linewidth]{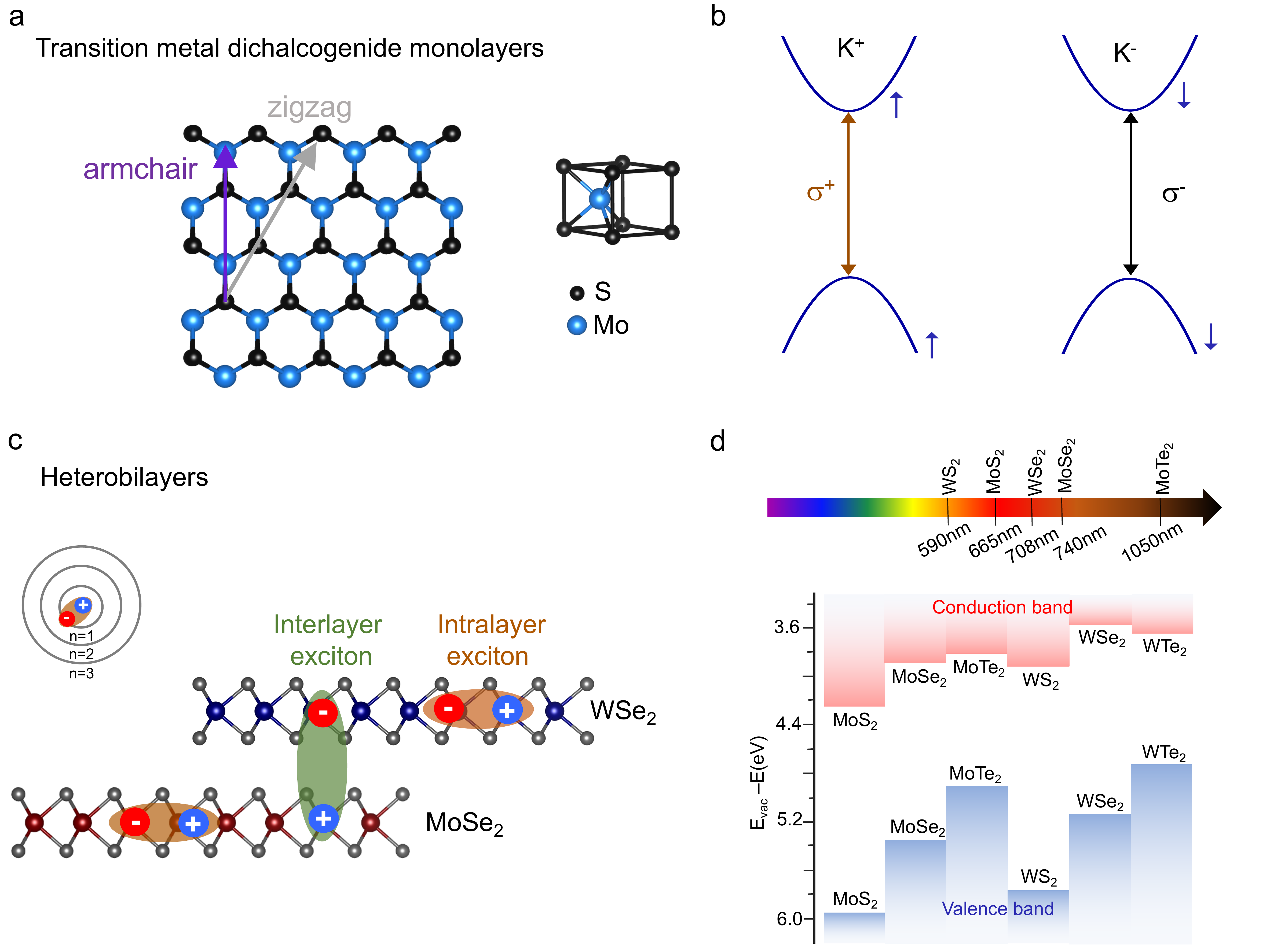} 
\caption{\textbf{Excitons in TMD monolayers and heterobilayers for optical spectroscopy}. \textbf{a,} Top view of MoS$_2$ monolayer hexagonal crystal structure, here transition metal atoms are shown in blue, the chalcogen atoms in black. The armchair and zigzag directions are indicated by purple and grey arrows. \textbf{b,} Schematic illustrations of valley selective optical transition which obey chiral selection rules addressed optically using $\sigma^+$ ($\sigma^-$) polarised light \cite{cao2012valley,xiao2012coupled,mak2012control,sallen2012robust,zeng2012valley}, see the details in Section \ref{polarization}. The direct bandgap is at the $K$-points of the Brillouin zone for TMD monolayers. \textbf{c,} Schematic of van der Waals heterobilayers. The individual monolayers show intralayer excitons (orange oval) a bound pair of an electron and hole which resides in a single layer. Interlayer excitons can be found in heterobilayers where the two charge carriers are located in different layers (green oval). Inset shows the hydrogen-like, exciton states which correspond to different energy levels, designated by their principal quantum number $n$. \textbf{d,} Band alignment between 2D semiconductor monolayers calculated in \cite{liu2016van}. Arrow shows the wavelength of the bandgap for different materials varying from the visible to near-infrared. }
\label{fig1}
\end{figure*}

\subsection{Optical properties of layered semiconductors : transition metal dichalcogenides}

\indent Interlayer van der Waals forces are considerably weaker than intralayer covalent bonding which makes a very large number of crystals exfoliable \cite{frindt1966single,mounet2018two}. TMD monolayers consist of a hexagonally oriented transition metal layer covalently bonded with top and bottom chalcogen layers \cite{dickinson1923crystal,wilson1969transition}, see FIG.~\ref{fig1}a. Multilayers are coupled by weak van der Waals forces and their symmetry is given by the stacking order \cite{van2019stacking,frondel1970molybdenite}, $i.e.$ the relative twist angle being 0$^{o}$ or 60$^{o}$. TMDs such as MoS$_2$ occur in different phases with semiconducting or metallic properties, here we focus on the semiconducting phase (2$H$).
The electronic and optical properties of TMDs change drastically as their thickness is reduced from bulk to atomic bilayer and monolayers. Bulk and few-layered TMDs exhibit an indirect bandgap, however, TMD monolayers show a direct bandgap. This gives rise to strong photoluminescence (PL) emission \cite{mak2010atomically,splendiani2010emerging} and also to a large, tunable absorption in monolayers \cite{tonndorf2013photoluminescence,moody2015intrinsic}. The light-matter interaction is dominated by excitons, Coulomb bound electron hole pairs, and this interaction is strongly enhanced when the incoming electromagnetic wave is resonant with the energy of the excitonic states \cite{wang2015giant}.\\
\indent The binding energies of excitons in TMD monolayers are of the order of several hundreds of meV. This is a consequence of the large electron and hole effective masses \cite{kormanyos2015k}, the reduced dielectric screening and the spatial confinement of the carriers. As a result, excitonic effects dominate the optical properties even at room temperature and beyond \cite{he2014tightly,chernikov2014exciton,ugeda2014giant,wang2018colloquium}.
A strong spin-orbit splitting of about 200~meV (for Mo-based) and 400~meV (W-based) appears in the valence bands at the $K$-point \cite{ramasubramaniam2012large,song2013transport,kormanyos2015k}. As a consequence, two separate interband optical transitions are observed in absorption, named A (transition from the upper valence band) and B (transition from the lower valence band) \cite{wilson1969transition}.\\
\indent Interestingly, the strong Coulomb interaction is extremely sensitive to the dielectric screening associated with a spatially inhomogeneous environment \cite{rytova2018screened,Keldysh} as discussed in Section \ref{spectroscopy}. 
This implies that the exciton transition energy, and to a larger extent, both the exciton binding energy and the free carrier bandgap can be tuned by engineering the local dielectric environment \citep{raja2017coulomb,waldecker2019rigid}. However, a possible undesirable consequence is that local dielectric fluctuations from disorder and impurities can result in strongly broadened optical transitions \cite{rhodes2019disorder,cadiz2017excitonic}.\\
\indent Optical transitions that are dipole-, momentum- and spin-allowed are referred to as 'bright' and the excitons recombine by emitting a photon. In layered materials excitons with different spatial orientations of the optical dipole, either in-plane or out-of the layer plane (see scheme in FIG.~\ref{fig:box}), participate in optical transitions in WSe$_2$ \cite{zhou2017probing} and InSe$_2$ \cite{brotons2019out}, for example. Taking into account also optical transitions that rely on phonon absorption or emission or spin-mixing of the different electronic states results in a large number of possible optical transitions ('bright' and 'dark') for a specific material.
The emission and detection efficiency can be optimised in applied magnetic fields and by selecting a particular light polarisation and propagation direction \cite{wang2017plane,robert2017fine,robert:2020measurement}. These  different types of optical transitions can be addressed selectively in the optical microscope set-up as detailed below. Mainly the \textit{interband} transitions between the valence band and conduction band are probed, but also transitions between excitonic states \cite{pollmann2015resonant} or \textit{intersubband} transitions in multilayers can be observed \cite{schmidt2018nano}.\\ 
\indent When the electron and hole reside within the same layer, we speak of intralayer excitons, as sketched in FIG.~\ref{fig1}c. Interlayer excitons can form in TMD heterobilayers due to the type-II (or staggered) band alignment \cite{liu2016van} with the photoexcited electrons and holes residing in different layers, see FIG.~\ref{fig1}d. These excitons are referred to as indirect in real space. The difference in the lattice constants between the monolayers forming the heterobilayer affects (i) the alignment of $K$-points which determines if the interlayer transition is direct or indirect in reciprocal space,  i.e. a phonon is needed in addition to a photon in the emission process \cite{kang2013band,rivera2015observation} and, (ii) formation of moir\'e effects/reconstructions \cite{van2014tailoring,sushko2019high,weston2019atomic,sung2020broken}. The latter leads to a periodic modulation of the electron and hole bandstructure depending on the difference between the lattice constants and/or twist angle, see FIG.~\ref{fig3}a. The depth and periodicity of the moir\'e potential can generate localised emitters (individual excitons) or collective excitations (trapping of excitons) \cite{yu2017moire}, with encouraging first reports all using optical spectroscopy \cite{shimazaki2020strongly,regan2020mott,wang2019evidence}. This approach for generating periodic, nanoscale potentials can certainly be extended to new types of van der Waals heterostructures.
\subsection{Layered semiconductor samples : monolayers and heterostuctures}
\label{fabrication}
\begin{figure*}[t]
\centering
\includegraphics[width=0.99\linewidth]{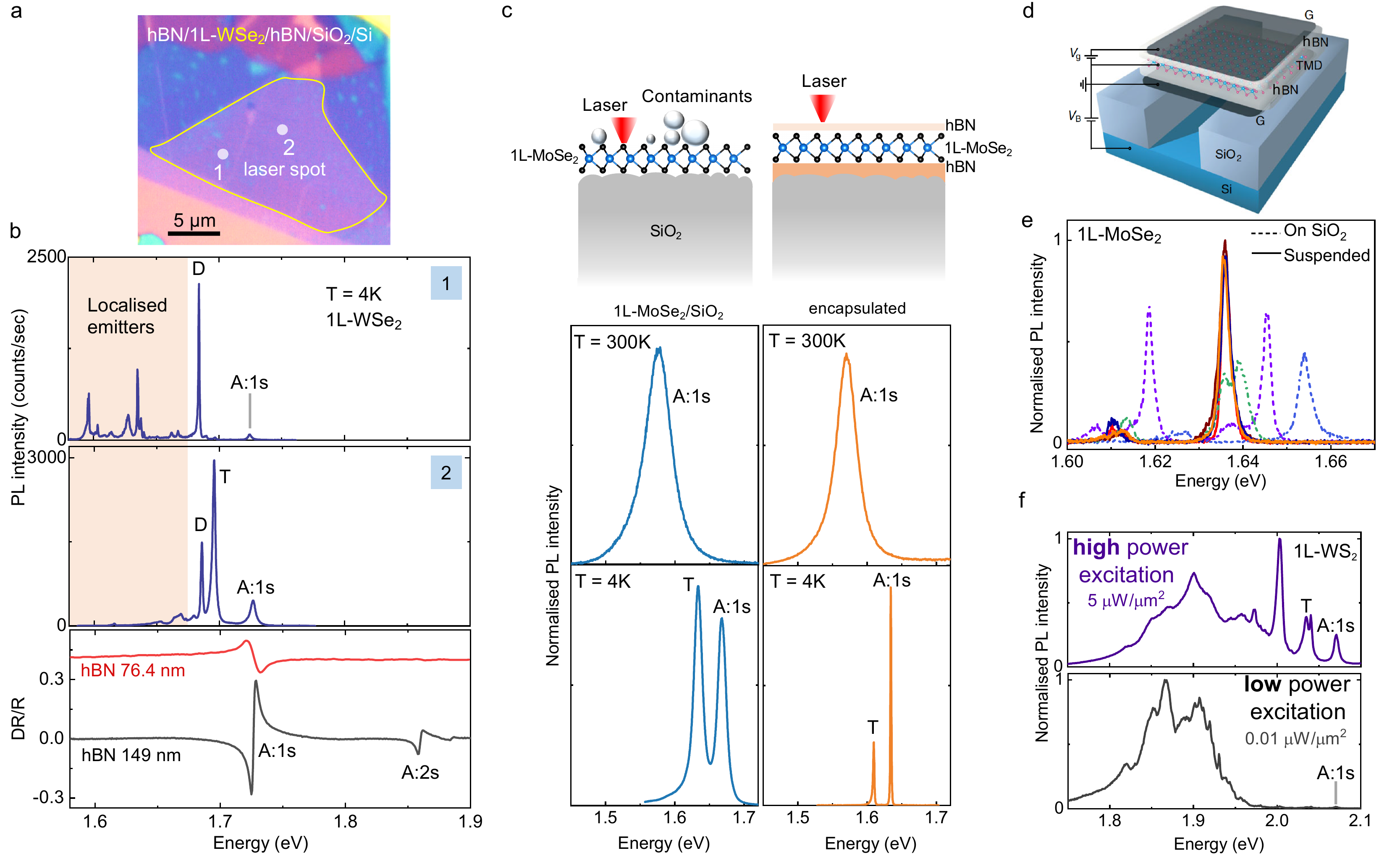} 
\caption{\textbf{Variation in photoluminescence (PL) response for different experimental conditions}. \textbf{a,} Optical micrograph of a typical van der Waals heterostructure sample containing a WSe$_2$ monolayer between hBN top and bottom encapsulation layer, sample homogeneity and sample imperfections (bubbles and wrinkles) can be seen. The homogeneous and non-homogeneous regions of the sample are labelled 1 and 2 in white, respectively. \textbf{b,} Typical PL spectra at T = 4~K recorded at two different locations 1 and 2 of the sample shown in \textbf{a}. The main exciton transitions (neutral excitons (A:1s), charged excitons (T) and spin- forbidden dark exciton (D) are quenched and localised emitters appear when PL is recorded on the bubbles or on wrinkles. However, strong PL emission corresponding to the main excitons from WSe$_2$ monolayers is recorded on the flat region of the sample. \textbf{c,} Sketch of MoSe$_2$ monolayer sample structures in different dielectric environment. The left panel is the sample structure of a non-encapsulated TMD monolayer on SiO$_2$ and the TMD monolayer encapsulated in hBN is on the right panel. Typical PL spectra recorded at room temperature and at cryogenic temperature on the encapsulated and the non-encapsulated sample are shown in the bottom panel. The linewidth reduces significantly for the encapsulated sample at T = 4~K compared to non-encapsulated sample. 
\textbf{d,} hBN encapsulated MoSe$_2$ layer suspended over a trench \cite{PhysRevLett.124.027401} and corresponding spectra are shown in \textbf{e}. 
The PL emission energy remains constant on suspended samples on different locations on the monolayer showing sample homogeneity \cite{PhysRevLett.124.027401}. 
 \textbf{f,} WS$_2$ monolayer encapsulated in hBN on a SiO$_2$/Si substrate, WS$_2$ layer plasma treated to generate optically active defects. Typical PL emission spectra for CW laser (532~nm) excitation at 0.01 $\mu$W and 5 $\mu$W  at T=4~K. The PL emission intensity of the main excitons are clearly visible at high power density, whereas these features are almost not detectable at low laser power density. PL spectra at low laser power density reveal that carriers can be trapped efficiently by defect sites and recombine by emitting photons at lower energy. Therefore at low laser power the PL emission of defects is considerably stronger than the free exciton emission.}
\label{fig2}
\end{figure*}

\begin{figure*}
\centering
\includegraphics[width=0.80\linewidth]{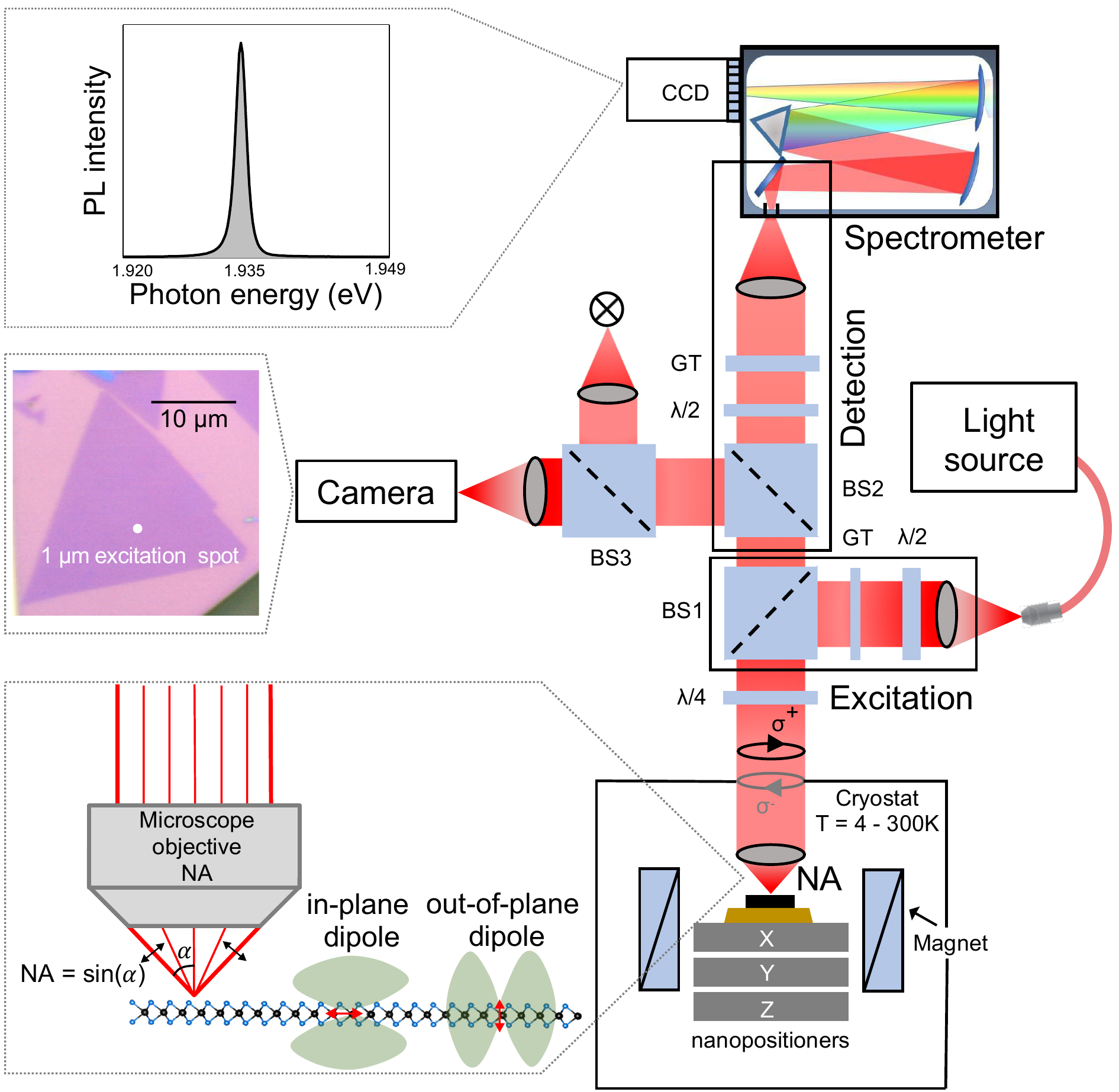} 
\caption{\textbf{Description of a typical microscope for optical spectroscopy}: Micro-photoluminescence and reflectivity set-ups have to fulfill several key criteria: High spatial resolution for sample mapping and accessing specific sample areas, high detection efficiency for example for quantum emitters in layered materials, combination of two detectors (one CCD/CMOS camera for sample imaging to guide the experiment and another camera to accumulate the emission signal as a function of wavelength). All this can be combined in a microscope, where excitation and detection pass through the same objective (epifluorescence geometry). The spatial resolution is given by the diameter $\Delta$ of the detection spot. This depends on the numerical aperture ($NA$) of the objective and the wavelength $\lambda$ through the Rayleigh criterion as $\Delta=\frac{0.61\lambda}{NA}$ \cite{hecht2002optics}, where $NA>0.8$ is common in commercial objectives. Excitons in layered materials can have an optical dipole either in-plane or out-of-plane, see inset representing emitters with in-plane and out-of-plane optical dipoles. 
Objectives with high $NA$ also collect part of the emission from an out-of-plane optical dipole.
A typical microscope set-up can be used to perform with similar equipment photoluminescence, reflectance, photoluminescence excitation (PLE), Raman scattering and second harmonic generation (SHG) spectroscopy experiments. Van der Waals heterostructures of different materials can be studied for temperatures T = 4 - 300 K. The samples are placed on nonmagnetic nanopositioners (travel distance several millimetres). An objective located inside a cryostat is more stable for long signal integration times. The other microscope components remain at room temperature. The microscope design has several modules: The excitation light propagates through an optical fibre via the lower horizontal arm. The upper horizontal arm consists of a camera and broad-band light source for sample imaging. The vertical arm is used for detection. The signal is collected by a spectrometer (here we show a Czerny-Turner geometry) with a diffraction grating coupled to a highly sensitive Si-CCD camera. Optionally an optical fibre can be used in the detection path. The fibre transports the signal from the microscope to the spectrometer entrance and also acts as a spatial pin-hole, making this a confocal arrangement for increased lateral (xy) and depth (z) resolution \cite{pawley2006handbook}. Polarisation analysis is a key motivation for studying layered materials such as TMDs for spintronics and valleytronics. Both linear polarisation and circular polarisation analysis can be achieved. For example, in the excitation part a Glan-Taylor (GT) polariser and a halfwave plate ($\lambda/2$) can be inserted. It is important to take into account the impact of the beamsplitters (BS) on the polarisation state. The quarter wave plate ($\lambda/4$) transforms the linear into circular polarisation and vice versa. Instead of a simple waveplate, a liquid crystal retarder (where a precise retardance of $\lambda / 2$, $\lambda / 4$ etc. can be adjusted by the applied bias) can optionally be used in the detection path. This avoids a macroscopic beam movement which can occur during rotation of a typical waveplate as the retardance/polarisation is changed. The performance of all optical set-up components is wavelength dependent, such as limited bandwidth of reflection coatings, chromatic aberrations of the objective or optical retardance for specific wavelengthd. Examples of optical components and polarisation analysis/rejection used in a microscope can be found in \cite{pawley2006handbook,kuhlmann2013dark,benelajla2020extreme}. }
\label{fig:box}
\end{figure*}

Widespread fabrication methods of monolayer samples on common SiO$_2$/Si substrates include mechanical exfoliation from bulk crystals (top-down) and bottom-up chemical vapor deposition (CVD). Promising results of high quality monolayers are also obtained by molecular beam epitaxy (MBE) \cite{dau2018beyond,pacuski2020narrow}.\\
\indent Exfoliated TMD crystals from high quality bulk show commonly defect densities of around $10^{12}$ $\text{cm}^{-2}$  \cite{rhodes2019disorder}, which is still considerably larger than in III-V semiconductor nanostructures. Currently the impact of defect concentration and type on the luminescence efficiency is widely investigated, with improvements reported for samples treated with super-acids \cite{amani2015near}. Exfoliation is widely used  because of simple handling even outside clean-room facilities and cost efficiency. However, there are several limitations: (a) the location of a single monolayer on the stamp/substrate is random and searching for a monolayer among flakes of different thickness is time consuming (b) the monolayers are relatively small with an average lateral dimension of tens of micrometer, and (c) the yield for finding monolayer per unit of surface area is low. \\
\indent As an alternative to exfoliation, CVD allows direct growth of monolayer material on a large surface area
\cite{lee2012synthesis,kobayashi2015growth,rhyee2016high,george2019controlled}. Here, the monolayer dimensions and the number of monolayers per unit of surface area are considerably larger than for exfoliation, see FIG.~\ref{fig:box}, with the limiting case of monolayer material covering the entire substrate surface. Recently, it has been demonstrated that lateral, as well as vertical TMD heterostructures can directly grow on flat or patterned SiO$_2$/Si substrates \cite{sahoo2018one,li2020general}. 
Detaching CVD-grown samples from the growth substrate to deterministically fabricate heterostructures is possible using, for instance, water assisted pick-up techniques \cite{jia2016large,paradisanos2020controlling}. \\
\indent For CVD-grown and exfoliated samples, the intrinsic quality can be evaluated by low-temperature optical spectroscopy. Emission from defects and inhomogeneous broadening in the optical spectra can be used as a diagnostic tool. In addition to the intrinsic layer quality, the impact of the underlying substrate can also be a limiting factor of the optical quality. This has been demonstrated by comparing low-temperature optical spectra between TMD monolayers (both CVD and exfoliated) on SiO$_2$ and encapsulated with hexagonal boron nitride (hBN) 
\cite{cadiz2017excitonic,raja2019dielectric,shree2019high}. It is possible to further improve the optical response by optimising the thickness of the top and bottom hBN in encapsulated TMD samples \cite{lien2015engineering,robert2018optical,fang:2019control}, see FIG.~\ref{fig2}b.\\
\indent Individual layers from high quality bulk crystals of different layered materials can be assembled into van der Waals heterostructures by using deterministic dry stamping \cite{Gomez:2014a} or direct pick-up \cite{purdie2018cleaning}. 
This allows controlled transfer of layers at precise locations on the substrate. An optical micrograph of a van der Waals heterostructure is shown in FIG.~\ref{fig2}a. During the transfer micro-bubbles (blisters) can appear due to trapped air, water or hydrocarbons \cite{purdie2018cleaning}. Agglomeration of the bubbles can be achieved by thermal annealing. This leaves clean, smooth areas with sharp interfaces \cite{doi:10.1021/acs.nanolett.7b01248}. In layered TMDs, the presence of bubbles or wrinkles can introduce defect emission due to strain and/or carrier localisation. This can lower the optical transition energy and lead to the appearance of localised emission, see FIG.~\ref{fig2}b and section \ref{spectroscopy}.\\
\indent Besides the intrinsic carrier density of a sample, charge impurities from a disordered substrate or adsorbates can introduce charge potential fluctuations, which strongly impact the optical properties. This is commonly observed in TMD monolayers on SiO$_2$/Si. As a result, transitions corresponding to charged exciton states (trions) can be detected in low temperature PL, shown in FIG.~\ref{fig2}b,c. Independent control of the carrier density in gated field-effect devices  
\cite{wang2017probing,PhysRevLett.124.027401} is thus crucial to study the optical properties  in the charge neutral, p- or n-doped regimes \cite{courtade2017charged,wang2017probing}. 


\section{Measuring optical absorption and luminescence in layered semiconductors}
\label{spectroscopy}
The fundamental optical transitions in TMDs lie in the energy range from $\approx$ 1.1~eV (monolayer MoTe$_2$) up to $\approx$ 2.1 eV (monolayer WS$_2$), see FIG.~\ref{fig1}d. Ferromagnetic semiconductors, such as CrBr$_3$ and CrI$_3$ cover similar transition energies \cite{bermudez1979spectroscopic,molinasnchez2019magnetooptical}. Interlayer excitons in heterostructures such as MoS$_2$/WSe$_2$ can reach emission wavelengths above 1100~nm ($<$1.1~eV), approaching the telecommunication bands \cite{karni2019infrared}. 
Black phosphorous is a layered semiconducting material with a direct bandgap that strongly varies with the number of layers and covers the visible (monolayer) to mid-infrared (bulk) spectral region \cite{ling2015renaissance}. The same evolution of band gap change versus thickness occurs in PtSe$_2$ \cite{ansari2019quantum}, but here the bandgap is indirect as for Si, so applications for detectors are possible.
At the opposite end of the spectrum layered hexagonal BN has a bandgap in the deep ultraviolet at 6~eV (200~nm) \cite{cassabois2016hexagonal}.
Below we describe the working principles of experiments to study absorption and emission of the optical transitions in layered semiconductors.

\subsection{Optical spectroscopy equipment}
\label{equipment}
A typical spectroscopy set-up contains a light source and several optical components to guide the excitation light to the sample and the signal to the spectrometer (monochromator) and detector. A charged-coupled device (CCD) or a complementary metal–oxide–semiconductor (CMOS) camera is also essential for the sample imaging. In this section we outline the characteristics of the main components and explain in FIG.~\ref{fig:box} the working principle of a versatile micro-spectroscopy set-up, widely used in commercial and also home-built systems. \\
\indent \textbf{Light sources.}- The main parameters for the laser excitation depend on the application and the sample's band structure: (i) the wavelength is selected in accordance to the investigated optical transitions (ii) the laser can be continuous-wave (CW) or pulsed. Pulsed lasers are more suitable for time-resolved experiments. For femtosecond (fs) or picosecond (ps) pulse duration the spectral width should be taken into account when investigating transitions close in energy. At the same time, the laser peak power should be calculated to avoid sample damage. High beam quality (aiming for perfect collimation, i.e. low $M^{2}$ factor) lasers are preferable for focusing the beam tightly to a diffraction limited spot, considering the small lateral dimensions of many exfoliated samples. For CW excitation, laser diodes can be typically selected between 375~nm and 2000~nm. For experiments requiring tunable wavelength excitation, convenient solutions for the 700-1000~nm range include Titanium Sapphire (Ti:sapphire) lasers \cite{spence199160}, which can be either pulsed (ps/fs) or continuous.
To cover the wavelength range between 500-700~nm and 1000-1600~nm, an optical parametric oscillator can be coupled to the Ti:sapphire laser combined with a doubling crystal. Dye lasers can also be used, where the choice of dye and its solvent allows for the selection of the emission range. Absorption or reflection measurements are performed using a broadband  white-light source to cover the full visible wavelength range. Often a simple halogen lamp suffices. When only one specific optical transition is investigated, a monochromatic LED or SLED with 10-20~nm spectral bandwidth can be used. This enables a good compromise between a small spot size and sufficient excitation power. Other solutions include laser driven light-sources or super continuum white lasers \cite{alfano2016supercontinuum} that allow to select a broad or narrow wavelength range for excitation, which makes them versatile for photoluminescence but also for broad band absorption experiments. \\
\indent \textbf{Optical components.}-  The optical components used in the set-up are selected for a specific wavelength range depending on both the excitation source and the emission wavelength. These include the polarisation components, microscope objectives, and lenses (ideally achromatic doublets) shown in FIG.~\ref{fig:box}. Homogeneous areas (flat surface, no charge fluctuations) in typical exfoliated sample have lateral dimensions down to a few micrometers. Therefore small excitation/detection spots close to the diffraction limit are crucial to record spectra with transition linewidths limited by the homogeneous, not inhomogeneous, broadening. A diffraction limited spot diameter can be achieved by using high numerical aperture ($NA$) objectives. \\
\indent \textbf{Detection.}- The final target is to detect the intensity as a function of wavelength of the light emitted from/scattered by the sample. The signal is focused onto  the entrance slit of a spectrometer. The collected signal is then dispersed by a monochromator, which can host different diffraction gratings, where a small (large) number of lines/mm allows studying a broad (narrow) spectral range. The signal can then be detected by a CCD or high quality CMOS chips. Alternatively, the monochromator can be left out and a simple combination of filters can be used in front of the detector. \\
\indent \textbf{Experimental conditions.}- Control of the ambient conditions is crucial. Many experiments are carried out at room temperature but low temperature experiments are necessary to access particular optical transitions. In simple bath cryostats the sample is kept in thermal contact with a liquid helium bath at T = 4~K, either via helium exchange gas or a cold finger. The main drawback is that the bath needs to be periodically refilled with liquid helium. Alternatively, closed-cycle cryostats liquify the helium gas using external compressors and allow continuous operation. The external compressor needs to be mechanically decoupled from the sample space to minimise vibrations. A piezo-based 3-axis stage with nanometer step-size is used to place a specific area of the layer of interest in the focal point of the objective. For high mechanical stability of the set-up, a low-temperature compatible microscope objective can be placed inside the cryostat, see FIG.~\ref{fig:box}. \\
\indent The sample holder can also be placed inside the bore of a superconducting coil to apply magnetic fields. In this case the nanopositioners, as well as the objective lens must be made of non-magnetic materials such as titanium and beryllium copper. The sample needs to be placed at the centre of the coil which limits in practice the sample size and also the optical access (beam diameter). The motivation for magneto-optics is manifold such as extracting the valley Zeeman splitting, and hence identifying the origin of new excitonic transitions including interlayer excitons in TMD homo- and heterobilayers, trilayers and bulk \cite{arora2017interlayer,leisgang2020giant}. Furthermore, it is possible to investigate valley polarisation dynamics and 'brightening' of otherwise spin forbidden-transitions \cite{zhang2017magnetic}. Attention should be given to undesired Faraday effects in certain optical components. Magnetic ions in the glass result in an undesired rotation of the linear polarisation in the presence of magnetic fields. This needs to be compensated by other polarisation control elements \cite{wang2016control}.

\subsection{Absorption spectroscopy}
\label{absorption}
Strictly speaking, measuring absorption ($A$) requires to measure both transmittance ($T$) and reflectance ($R$) where $A = 1-R-T$. Transmission measurements require a transparent substrate and a detection path different from the excitation path, for example a separate microscope objective on each side of the sample or alternatively  one objective combined with a fibre on the other side of the sample \cite{goryca2019revealing}. In practice often reflectivity is measured as it is the simplest experiment for samples on substrates like SiO$_2$/Si that are not transparent. To get a quantity independent of the optical response of the set-up, one generally measures the reflectivity contrast defined as  $(R_\text{sam}-R_\text{sub})/R_\text{sub}$, where $R_\text{sam}$ is the intensity reflection coefficient of the sample with the TMD layer and $R_{sub}$ comes from the same structure without the TMD layer. The spectra obtained this way are for brevity commonly referred to as absorption in the literature. 
\\
\indent The optical properties of a material can be seen in a simple classical picture as the interaction between light (electromagnetic radiation), and various types of oscillators \cite{klingshirn2012semiconductor}. In TMD monolayers, the dominating oscillators are exciton resonances. Therefore, in reflectivity different exciton resonances are accessible up to room temperature as they possess strong oscillator strength and high density of states (DOS). This allows to observe the Rydberg series of the A-exciton: 1$s$, 2$s$, 3$s$...(see FIG.~ \ref{fig2}b bottom panel), thus giving a measure of the exciton binding energy and the single particle bandgap \cite{chernikov2014exciton,goryca2019revealing}. Other optical transitions related to defect states in the gap or other exciton complexes which posses weaker oscillator strength and/or comparatively lower density of states are difficult to trace in absorption, although they might appear in photoluminescence emission, as discussed below. \\
\indent \textbf{Monolayers.}- The energy of the A-exciton transition in TMD monolayers is given by the difference of the single particle bandgap (of unbound electrons and holes) and the exciton binding energy. Engineering the dielectric environment (and hence all energy scales linked to the Coulomb interaction) results in significant changes in the single particle bandgap \cite{waldecker2019rigid} and the exciton binding energy of TMD monolayers. But the shift in the global A-exciton transition energy is rather small, as changes in single particle bandgap and the exciton binding energies partially compensate each other. Compare FIG.~\ref{fig2}c for monolayer MoSe$_2$ transition energies with and without hBN encapsulation, which are very close in value. However, the linewidth in absorption is significantly impacted by dielectric disorder \cite{raja2019dielectric}. For instance, bubbles, wrinkles, polymer residues and hBN have different dielectric constants. Therefore, a non-uniform dielectric environment affects the energy of the exciton transitions and the overall shape of reflectivity spectra. Uniform dielectric slabs such as thick (tens to hundreds of nm) hBN layers can be exploited to steer the absorption. The visibility of exciton resonances in absorption is mainly influenced by the thickness of hBN and SiO$_2$, see FIG.~\ref{fig2}b. This is due to thin-film interference effects, the bottom hBN thickness determines how far the monolayer is from the Si/SiO$_2$ interface, which acts as a mirror. The choice of hBN thickness of the heterostructure can be optimised using a transfer matrix approach to increase the visibility of the targeted transitions \cite{robert2018optical}. In this process, the particular energy of the excitonic resonance of the layered semiconductor should be taken into account \cite{robert2018optical}. Recent results in TMD materials placed in front of mirrors show a modulation in the absorption strength of up to 100\% due to interference/cavity effects \cite{horng2019perfect,scuri2018large,back2018realization}. The strong influence of the dielectric environment on the light-matter interaction of atomically thin semiconducting membranes motivates a great potential for sensing applications including novel device architectures with precisely tunable optical properties. \\
\indent \textbf{Multilayers.}- 
In general the nature (direct or indirect) and the energy of the bandgap evolves as a function of the layer number for a given material. 
In addition, the absorption of layered semiconductors such as black phosphorous \cite{ ling2015renaissance} and ReSe$_2$ \cite{ho1998absorption, zhang2016tunable} reveals information on the crystal structure of these particular materials as it is highly anisotropic in the layer plane, as a direct result of a highly anisotropic lattice structure. 
In TMD multi-layers and even bulk strong excitonic features are reported even at room-temperature in early studies \cite{wilson1969transition}. In addition to these features typically attributed to the intralayer A- and B-exciton, more recently the observation of interlayer excitons (formed by carriers in 2 adjacent layers) has been reported in absorption of bulk samples \cite{arora2017interlayer,horng2018observation,arora2018valley}. Absorption of interlayer excitons is also reported in homobilayers and homotrilayers of MoS$_2$ \cite{Gerber:2019,slobodeniuk2019fine}. In these systems the transition energy of the absorption can be tuned through the application of an electric field perpendicular to the layers (Stark shift) over 120~meV and the interaction between interlayer and intralayer excitons can be investigated \cite{leisgang2020giant,sung2020broken,lorchat2020dipolar}.

\begin{figure*}
\centering
\includegraphics[width=169mm]{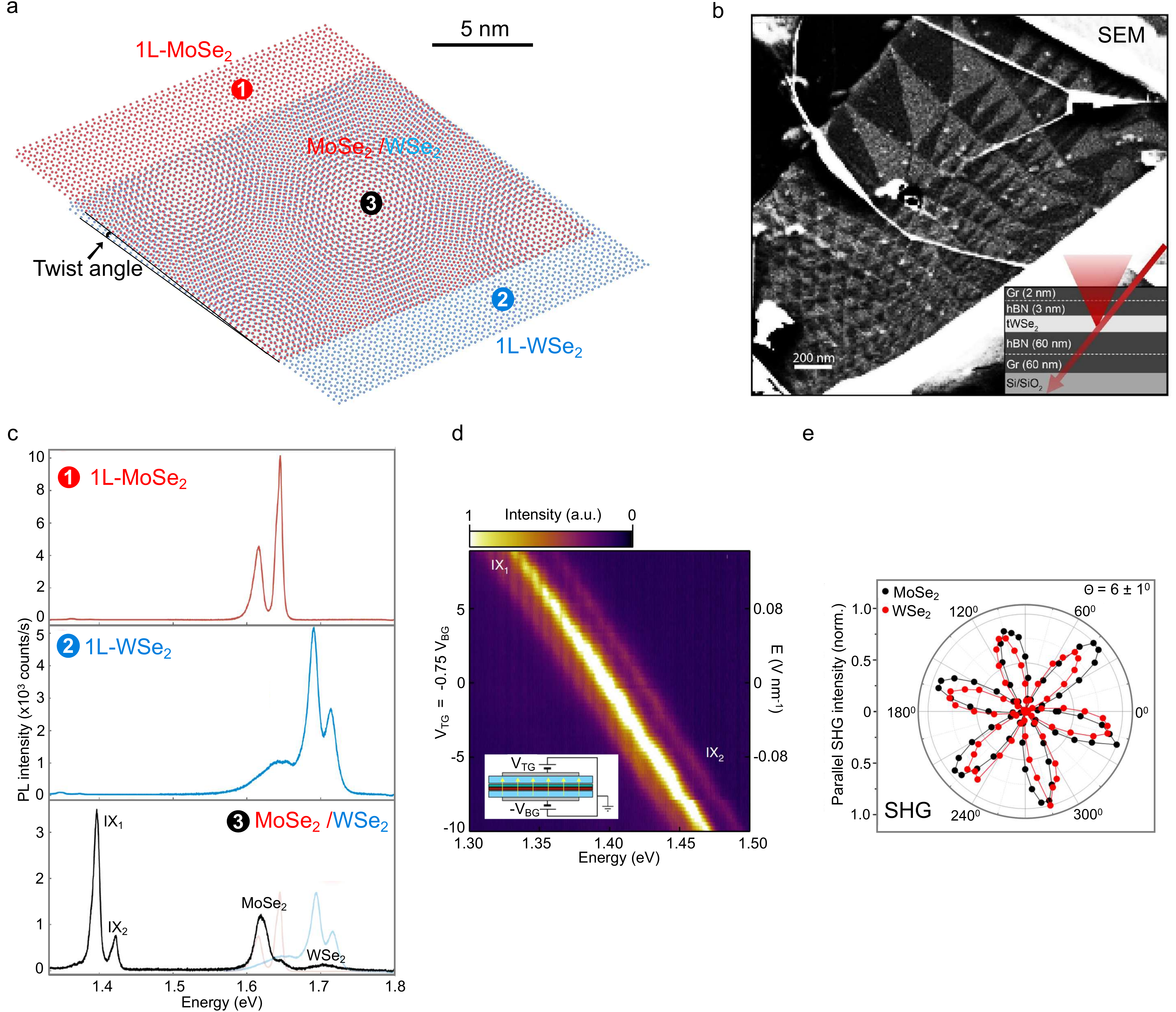} 
\caption{\textbf{Moiré interlayer excitons in heterobilayers.} \textbf{a,} Two different monolayer materials stacked vertically that display a moir\'e pattern due to slight lattice mismatch and twist angle. Different local atomic alignments in the heterostructure show different optical properties \cite{yu2017moire,shimazaki2019moir,ciarrocchi2019polarization,shimazaki2020strongly}.  Three spots labelled 1, 2 and 3 are chosen for PL measurements on MoSe$_2$ monolayer, WSe$_2$ monolayer and MoSe$_2$/WSe$_2$ heterostructure, respectively, and corresponding spectra are shown in \textbf{c}. \textbf{b,}  SEM imaging of hBN-encapsulated twisted WSe$_2$ bilayers with a spatially varying reconstruction pattern due to the interaction between the respective layers after stacking \cite{sushko2019high}. \textbf{c,} Example of photoluminescence spectrum from an hBN-encapsulated MoSe$_2$ monolayer (solid red curve), MoSe$_2$/WSe$_2$ heterostructure (solid black curve) and WSe$_2$ monolayer (solid blue curve) \cite{ciarrocchi2019polarization}. Intralayer exciton emission is observed from the MoSe$_2$ and WSe$_2$ monolayers. The interlayer exciton emission (IX$_1$ and IX$_2$)  appear in energy below the intralayer resonances from the heterostructure. \textbf{d,} Electrical control of interlayer excitons in MoSe$_2$/WSe$_2$ heterobilayer. Colormap of PL emission spectra as a function of applied gate voltages $V_{TG}$ and $V_{BG}$ when sweeping at constant doping. Stark shift of the interlayer excitons in applied electric fields perpendicular to the layers showing out-of-plane electric dipole \cite{ciarrocchi2019polarization}. \textbf{e,} Polarisation-resolved second harmonic generation (SHG) intensity of the individual monolayers from a different publication compared to panels \textbf{c,d}, indicating the armchair directions of the monolayers which determine the twist angle ($\theta$) between the WSe$_2$ and MoSe$_2$ layers \cite{nagler2017giant}.}

\label{fig3}
\end{figure*}
\subsection{Photoluminescence spectroscopy}
Luminescence experiments are widely used for studying the macroscopic optical properties of materials as well as their microscopic electronic excitation, for the evaluation of crystalline quality (presence of defects) and for testing novel optoelectronic devices \cite{pelant2012luminescence, haunschild2010quality}. Luminescence is defined as a surplus of the electromagnetic radiation (light) emitted by a solid, in addition to its equilibrium radiation described by Planck’s law. This surplus energy is transformed into detectable luminescence radiation. During the process of luminescence the electrons are excited to higher energy states (by a \textit{light}-source in the case of \textit{photo}-luminescence). Subsequently the carriers relax in energy for example through phonon emission, followed by photon emission. The succession of all involved relaxation and recombination events lasts a relatively long time, which is a main difference compared to other types of so-called secondary radiation: reflected light and scattered light (for example Raman). Once the material is excited with a light pulse, the luminescence continues to decay for some time and can be recorded in time-resolved photoluminescence \cite{pelant2012luminescence}, see a review on time-resolved spectroscopy \cite{balocchi2012time} for technical details. These experiments  give information on carrier relaxation and recombination times. In addition to time-resolved photoluminescence (an incoherent technique),  also important information can be gleaned from coherent spectroscopy such as four-wave mixing and two-color pump-probe experiments \cite{jakubczyk2016radiatively,moody2015intrinsic,hao2016coherent}. \\
\indent In absorption spectra optical transitions with large oscillator strength and high density of states dominate. In contrast, the emission spectrum given in photoluminescence experiments can be dominated by other transitions as these experiments probe the population of a state. Typically, optical transitions at lower energies are detected in PL as carriers relax towards these lower energy states before radiative recombination. It is therefore instructive to compare emission with absorption spectra (measured in reflection geometry) as in FIG.~\ref{fig2}b. For the material investigated, monolayer WSe$_2$, the main bright exciton transition (A:1s) that dominates in absorption is not generally the strongest feature in PL emission, as electron-hole pairs can relax towards states lying at lower energy to recombine \cite{wang2017plane}. \\
\indent \textbf{Sample temperature.}- At room temperature, the main transitions such as the A- and B-excitons in TMDs will be broadened due to scattering with phonons, compare spectra at T=4~K and 300~K in FIG.~\ref{fig2}c. Carriers or excitons are mobile at high temperature and defect potentials in the lattice with only shallow confinement energy will not act as efficient trapping sites. At low temperature phonon absorption is reduced and the linewidth reveals the sample quality (inhomogeneous broadening versus homogeneous broadening \cite{fang:2019control}). This is illustrated in FIG.~\ref{fig2}c. Spectral lineshape and main emission energy can change compared to high temperature. Carriers can get trapped at sufficiently deep defect potentials. Also excitonic complexes with lower binding energy, such as trions, are now stable and contribute to the PL signal, see FIG.~\ref{fig2}b,c,f for three different materials \cite{moody2015intrinsic,cadiz2017excitonic,Robert:2017,shree2018observation}.\\
\indent \textbf{Excitation power.}- Laser power plays an important role as it controls the number of photoexcited carriers. Strictly speaking, the power density (defined as the average power per unit area) is directly related to the photogenerated exciton density in layered semiconductors. It is inversely proportional to the square of the focused beam radius, thus it is directly related to the numerical aperture of the objective lens and laser wavelength. Let us take an example of a sample with a finite value of defect density: for a tightly focused beam (spot size $\approx$ 1~$\mu$m), at low laser power (typically hundreds of $n$W) all carriers can be trapped at defect sites and the PL signal of free excitons is not visible, as in FIG.~\ref{fig2}f, lower panel. Increasing laser power (few $\mu$W) fills all defect sites and free exciton PL can be measured in addition to defect emission, see FIG.~\ref{fig2}f upper panel. Further increase in power (tens to hundreds $\mu$W) will lead to such a high concentration of excitons that exciton-exciton interactions start to play a role. At high excitation density very different phenomena such as biexciton formation and exciton-exciton annihilation can be studied depending on the material \cite{nagler2018zeeman,sun2014observation,barbone2018charge,paradisanos2017room}. Understanding emission for high exciton (carrier) concentrations is crucial for applications such as lasing and for investigating collective states such as condensates \cite{wang2019evidence,sigl2020condensation}.\\
\indent \textbf{Sample quality and dielectric environment.}- Two monolayer samples exfoliated from the same quality bulk crystal but placed on different substrates can exhibit very different PL characteristics, see FIG.~\ref{fig2}c for a comparison of MoSe$_2$ on SiO$_2$ and on hBN, respectively. 
The total line broadening comes both from homogeneous contributions given by $\approx$ 1~ps lifetime (below 1 meV linewidth) and inhomogenous contribution from sample imperfections (impurities, defects, interface, substrate, etc) \cite{jakubczyk2016radiatively}. 
The key role of high quality hBN buffers with low defect density is to provide atomic flatness \cite{Taniguchi:2007a} for monolayer deposition and a very clean, homogeneous dielectric environment. The hBN bulk bandgap of  6~eV \cite{cassabois2016hexagonal} is high enough to use it as an essentially transparent encapsulation layer for many materials, providing an ideal environment to address intrinsic properties of 2D-TMDs and preserving good optical quality of air-sensitive materials such as CrI$_3$ or black phosphorous. As the inhomogenous broadening is largely suppressed in high quality samples of TMD monolayers, the linewidth starts to be an indication of the exciton lifetime (homogeneous broadening) and can be tuned by carefully choosing the encapsulating hBN thickness - placing the monolayer on a node or anti-node of the electromagnetic field in the multilayer structure \citep{fang:2019control}. \\
\indent PL emission is sensitive to the presence of wrinkles or bubbles which induce strain and localisation potentials in encapsulated monolayers. Small strain ($1\%$) in monolayer induces a rather significant (about 50 meV) shift in bandgap energy \cite{zhu2013strain} which can explain shifts in absolute emission energy from sample to sample and for different areas of the same sample. Therefore, in practice PL emission intensity as well as linewidth varies on different location of TMD monolayers, see FIG.~\ref{fig2}b when the detection spot (1 $\mu$m diameter) is scanned across a WSe$_2$ monolayer. Remarkably, exciton emission energy that does \textit{not} vary as a function of the detection spot position has been observed in hBN-encapsulated MoSe$_2$ monolayers, suspended over a trench, see FIG.~\ref{fig2}d,e \cite{PhysRevLett.124.027401}, indicating a stable, inert environment.\\
\indent The general study of the dielectric environment, surface quality, flatness, charging events and their impact on optical properties is very important also for other nanostructures such as carbon nano-tubes \cite{berger2009optical,hirana2010strong,ai2011suppression,noe2018environmental,raynaud2019superlocalization} and layered perovskites \cite{blancon2018scaling} which show strong excitonic effects. \\
\indent\textbf{Interference effects.}- Layered materials are usually placed on a substrate such as Si with an SiO$_2$ layer of typically 85~nm. As discussed earlier for white light absorption experiments, optical interference will also be important for the laser excitation beam and the PL emission as a function of the SiO$_2$ thickness and possibly the encapsulation layer thickness. Reversely, for constant SiO$_2$ thickness absorption and emission intensities and directivity will depend on the wavelength, as detailed in \cite{lien2015engineering,zhang2015interference}. In practice the thickness of the SiO$_2$ layer on top of Si is chosen to maximise optical contrast of monolayers already during sample fabrication, as discussed in detail for graphene on SiO$_2$ \cite{roddaro2007optical}. \\
\indent\textbf{Emission dynamics.}- The detected luminescence signal from a layered semiconductor is the result of an intricate interplay between radiative and non-radiative energy relaxation. Time-resolved PL can be performed using a pulsed laser excitation and measuring the recombination (emission) time. In clean TMD samples the strong exciton oscillator strength leads to an intrinsic radiative lifetime of the order of 1~ps at low temperature \cite{lagarde2014carrier}. Localised emitters recombine considerably slower and also dark excitons have a lifetime up to 3 orders of magnitude longer. In the time domain, low temperature measurements on high quality samples allow spectrally isolating each transition (either with bandpass filters or with a spectrometer) and then studying the emission dynamics of each optical transition separately \cite{Robert:2016a,Robert:2017,fang:2019control}.\\
\indent \textbf{Optical dipole orientation.}- In TMD monolayers the main optical transitions have an in-plane optical dipole, they therefore emit light normal to the monolayer plane. However, in addition to these bright (spin-allowed) transitions also excitons that have an out-of-plane optical dipole emit light \cite{wang2017plane}. Out-of-plane dipole transitions are also prominent in InSe \cite{bandurin2017high,brotons2019out}. Due to the small sample dimensions most experiments are carried out in a  microscope using an objective with high ($>$0.8) $NA$, see inset on microscope objective in FIG.~\ref{fig:box}. As a result, PL emission containing out-of-plane and also in-plane components in the monolayer are detected. In WSe$_2$ and WS$_2$ monolayers, dark excitons are prominent, see FIG.~\ref{fig2}b and lead to exotic, higher order complexes (such as biexcitons made up of a dark and a bright exciton and so-called dark trions \cite{barbone2018charge,PhysRevLett.124.196802}), that can be identified by monitoring the orientation of the exciton dipole. The role of out-of-plane dipole emission is also investigated for quantum emitters in WSe$_2$ \cite{luo2020exciton}. Brightening of (spin-) dark states due to an increased mixing of the spin-states in WSe$_2$, WS$_2$, MoSe$_2$ and MoS$_2$ monolayers can be observed in low-temperature magneto-PL experiments by applying strong (ideally several tens of T) in-plane magnetic fields \cite{robert2017fine,lu2019magnetic,robert:2020measurement}.\\
\indent \textbf{Multilayers.}- PL spectroscopy is useful also in TMD heterobilayers with type II alignement for the examination of spatially indirect interlayer exciton (see FIGs.~\ref{fig1}c and \ref{fig3}c) with large binding energies ($\approx$ 150 meV) \cite{rivera2015observation,rivera2018interlayer,jauregui2019electrical}. A long period moir\'e pattern (see sketch in FIG.~\ref{fig3}a) offers new directions to explore and control exciton arrays in twisted TMD heterobilayers from potentials that trap individual excitons to the formation of minibands. This allows physics related to the Mott-insulator \cite{shimazaki2020strongly}, for potential applications in quantum optoelectronic devices \cite{andersen2019moir,sung2020broken}. Some key characteristics of interlayer excitons include a long-lifetime (ns), a wide transition energy tunability that ranges over several hundreds of meV via applied electric fields, see FIG.~\ref{fig3}d and a characteristic Zeeman splitting when compared to intralayer excitons \cite{arora2018valley}. 
\subsection{Photoluminescence excitation spectroscopy}
In photoluminescence excitation spectroscopy (PLE) the PL emission intensity for a chosen energy is recorded for different photon excitation energies. Tunable lasers or powerful white light sources are used as an excitation source. The linewidth and tuning step of the source will determine the spectral resolution of the PLE experiment. The measured PL intensity will depend on two factors (i) the absorption strength at the excitation energy and (ii) the efficiency of energy relaxation followed by radiative recombination (in competition with non-radiative channels). This combined dependence on both absorption and energy relaxation (often through phonon emission) make PLE spectroscopy a very interesting tool for several investigations :  \\
\indent \textbf{Interlayer excitons.}- A PL signal enhancement of interlayer excitons is observed when the laser excitation energy is resonant with intralayer states in one of the layers, confirming that interlayer excitons form via charge transfer processes between the layers \cite{rivera2015observation,hong2014ultrafast,ciarrocchi2019polarization}, see FIGs.~\ref{fig1}d and \ref{fig3}c. In general electronic coupling or charge transfer between layers can be investigated by tuning a laser in resonance with an electronic transition in one layer and monitoring PL emission at an energy corresponding to the adjacent layer or to the heterostructure.\\
\indent \textbf{Measuring excited exciton states.}- 
PLE can be also used to establish a link between optical transitions with similar microscopic origin within the same monolayer. In MoS$_2$ monolayers, B-exciton states energetically overlap with the excited A-exciton states (A:2$s$, A:3$s$..,). PLE spectroscopy allows to distinguish the excited states by collecting the emission intensity of the ground state, A:1$s$, as a function of the excitation laser energy, scanned over the energy of A:2$s$, A:3$s$, etc. Besides states with $s$-symmetry, also $p$-states can be examined. To access $p$-states, two-photon absorption processes are necessary and therefore the laser energy needs to be tuned to half of the transition energy \cite{cassabois2016hexagonal}. The identification of high-excited exciton states in one and two-photon-PLE is a powerful method to evaluate the impact of different dielectric environments on the energy evolution of the exciton states. Furthermore, it is possible to extract the exciton binding energies \cite{hill2015observation} and investigate predictions of splittings of the $p$-exciton states \cite{srivastava2015signatures}. However, one should note that the crystal symmetry or disorder effects can mix $s$ and $p$ exciton states \cite{glazov2017intrinsic,berghauser2016optical}. \\
\indent \textbf{Identification of dominant phonon modes for energy relaxation.}- In addition to key information on absorption, PLE is used to identify efficient relaxation channels. In PLE experiments on MoSe$_2$ monolayers a periodic oscillation in energy is observed over an energy range without any expected exciton resonance (roughly constant absorption \cite{wang2015exciton}). These maxima are all equally spaced in energy by longitudinal acoustic phonons at the M point of the Brillouin zone, LA(M), revealing the efficient energy relaxation of excitons through emission of LA(M) phonons \cite{chow2017phonon,shree2018observation}. This experimental observation was possible due to the spectrally narrow excitation source, resolving fine separations between different peaks related to phonon emission \cite{shree2018observation,PhysRevB.93.155407}.   

\section{Accessing spin-valley polarisation in optical spectroscopy}
\label{polarization}
The symmetry of the electronic states in monolayers and multilayer crystals governs the optical selection rules for light polarisation in emission and absorption \cite{xiao2012coupled,yu2017moire}, see FIG.~\ref{fig1}b, as studied since several decades for semiconductor nanostructures \cite{dyakonov2017spin}. For polarisation analysis, linear polarisers and waveplates can be inserted in the detection and excitation path of the set-up, see FIG.~\ref{fig:box} for practical details. Exciting a system with polarised light can address a specific spin or valley state, see FIG.~\ref{fig1}b. The emitted light gives information on spin and valley dynamics in time-integrated PL experiments. The circular polarisation in time-integrated experiments depends on the exact ratio of PL emission time $\tau_{PL}$ versus depolarisation time $(\tau_{depol})$ as
$P_c = P_0/(1 + \frac{\tau_{PL}}{\tau_{depol}})$, where $P_0$ is the initially generated polarisation which could depend on the excitation energy \cite{kioseoglou2012valley,tornatzky2018resonance}. \\
\indent \textbf{TMD monolayers.}- The interband transitions in monolayer materials such as MoS$_2$ are governed by chiral selection rules as optical transitions in the $K^+$ ($K^-$) valley are $\sigma^+$ ($\sigma^-$) polarised. For neutral, bright excitons the intrinsic lifetime is of the order of 1~ps, so from time-integrated PL experiments that report $P_c$ values in the order of 50\% one can infer that $\tau_{depol}$ is at least of this order of magnitude. In practice more sophisticated pump-probe measurements reveal very short valley lifetimes for neutral excitons \cite{jakubczyk2016radiatively}. The ratio $\frac{\tau_{PL}}{\tau_{depol}}$ can be tuned by placing monolayers in optical microcavities \cite{dufferwiel2017valley, dufferwiel2018valley}. Longer valley lifetimes in monolayers are reported for resident carriers \cite{dey2017gate} not excitons, measured with pump-probe techniques such as Kerr rotation employed for probing polarisation in semiconducting or metallic nano-structures \cite{mak2019probing}.\\
\indent \textbf{TMD heterobilayers.}-  Optical spectroscopy can be used to probe the local atomic registry, i.e. how metal and chalcogen atoms are aligned in the top with respect to the bottom layer \cite{yu2017moire,seyler2019signatures,ciarrocchi2019polarization}. Here information can be gleaned on the formation of nano-scale, periodic moir\'e potentials, see sample sketch in FIG.~\ref{fig3}a. But similar to bilayer graphene, reconstruction can occur when two TMD layers are brought in contact, which can be visualised using imaging techniques such as transmission electron microscopy (TEM) or scanning electron microscopy (SEM). Recently Sushko et al. \cite{sushko2019high} reported SEM imaging of hBN-encapsulated twisted WSe$_2$ bilayers showing that a spatially varying reconstruction pattern develops due to the interaction between the respective layers after stacking, see FIG.~\ref{fig3}b. Polarisation selection rules probed in PL also carry information on different stackings (H-type or R-type for 60$^{\circ}$ or 0$^{\circ}$ twist angle, respectively) \cite{yu2018brightened}. Therefore, polarisation-resolved optical spectroscopy together with direct atomic-resolution imaging of the lattice is a very powerful combination for analysing the formation of moir\'e potentials \cite{andersen2019moir}. The PL experiment samples over a spot diameter of 1 $\mu$m, whereas moir\'e potentials can occur with a periodicity of nanometers (see FIG.~\ref{fig3}a), which leads to averaging effects. The intrinsic lifetime of interlayer excitons is of the order of ns at low temperature and not ps as in monolayers, which allows for imaging exciton and polarisation spatial diffusion in PL maps \cite{rivera2016valley, unuchek2019valley}. The physics of both intralayer and interlayer excitons can be accessed in the monolayer and bilayer regions of the same sample, as in FIG.~\ref{fig3}a and c. \\
\indent \textbf{Experiments in applied magnetic fields.}- The circular polarisation can be manipulated by applying external magnetic fields \cite{molas2019probing,srivastava2015valley}. Interesting examples are heterobilayers, where a giant Zeeman splitting of 26~meV at B = 30~T for interlayer excitons induces near-unity valley polarisation measured in PL emission \cite{nagler2017giant}. In monolayer MoSe$_2$ a field of 7~T results in near unity polarisation of electrons probed in absorption and emission \cite{PhysRevLett.118.237404}.

\section{Optical techniques for accessing crystal quality and orientation}
\label{orientation}
\indent \textbf{Raman spectroscopy} is based on the analysis of laser light scattered by a material. During this process the crystal typically absorbs (or emits) energy in the form of lattice vibrations - phonons. The analysis of the scattered light's energy and polarisation reveals information on the crystal symmetry and quality, doping and where applicable alloying and stacking in multilayers.  A typical PL set-up (shown in FIG.~\ref{fig:box}) can be conveniently adapted to collect the Raman spectra by selecting a suitable set of filters according to the wavelength of the excitation laser. Typically, filters for Raman spectroscopy reject the excitation laser with a cut-off frequency of few tens of $cm^{-1}$ ($\approx$ 10~meV from the single mode energy). 
The selection of laser wavelength $\lambda$ has an important impact on the spectral sensitivity as the intensity of the Raman signal is proportional to $\lambda^{-4}$. 
Key parameters of the excitation laser include spectral linewidth ($\leq$1~MHz), frequency and power stability, spectral purity ($\geq$65~dB side-mode suppression ratio), beam quality (close to Gaussian) and output power. 
The excitation wavelength in the Raman scattering of TMDs is also important because of the presence of excitonic states. When the photon energy matches the transition energy of a real state it gives rise to a strong signal enhancement and appearance of new features, associated to symmetry dependent electron-phonon interactions (resonant Raman scattering) \cite{PhysRevLett.114.136403,scheuschner2015interlayer}.\\
\indent Information on the structural phase and composition of materials can be obtained by means of Raman spectroscopy \cite{oliver2017structural}. 
This allows for example investigating currently debated link between ferromagnetic ordering and structural phase transitions in CrI$_3$ \cite{ubrig2019low} as a function of temperature. Raman spectroscopy can reveal anisotropy in the crystal structure of, for example, ReSe$_2$, which can be directly linked to measurements of anisotropic optical absorption in the same material \cite{wolverson2014raman}.
Electron-phonon interactions can have significant effects on the Raman frequencies. As a result, doping effects can be effectively monitored in TMD monolayers for example with the out-of-plane phonon, $A'_{1}$, due to its strong electron-phonon coupling \cite{chakraborty2012symmetry,bertolazzi2017engineering}. It is also possible to extract quantitative information about the presence of uniaxial strain since the in-plane phonon energy, $E'$, decreases with applying tensile strain and a splitting occurs (degeneracy is lifted) \cite{conley2013bandgap}. An estimation of the monolayer crystal quality and presence of defects can be realised due to the activation of defect-induced zone-edge phonon modes, such as the LA(M) \cite{mignuzzi2015effect}. Furthermore, disorder and interference effects originating from the substrate impact the intensity and spectral shape of the optical phonons in the monolayer \cite{buscema2014effect}. A powerful and reliable means to determine the number of TMD layers with atomic-level precision is to measure the energy difference between the two main vibrational modes ($E$ and $A$ phonons), affected by interlayer interactions \cite{lee2010anomalous}. Apart from the high-frequency ($\geq$ 80~$cm^{-1}$) spectral range, the number of layers can be identified by collecting optical signatures of the rigid layer vibrations (breathing and shear modes) in the ultra-low frequency range \cite{zhang2013raman}. 
In this case, important information on the interlayer interaction and determination of the stacking order in multilayers can be obtained \cite{van2019stacking}. In TMD multilayers resonant Raman spectroscopy can also provide a fingerprint of the extension of excitons over several layers, as otherwise symmetry forbidden modes are activated for the so-called C-exciton region \cite{scheuschner2015interlayer} in energy above the A- and B-exciton. Raman spectroscopy can also be used to investigate the competition between formation of periodic moir\'e potentials and local reconstruction (compare FIGs.~\ref{fig3}a and b) in artificially stacked WSe$_2$/MoSe$_2$ and bilayer MoS$_2$ as a function of twist angle \cite{holler2020lowfrequency,debnath2020evolution}.\\
\indent \textbf{Second-harmonic generation} (SHG) is a nonlinear optical process that converts two photons of the same frequency into one photon of twice the orignal frequency. It is a powerful technique to analyse the orientation and symmetry properties of 2D materials. For SHG experiments, the optical set-up is typically coupled to a pulsed laser that is capable of generating sufficient peak power for this nonlinear optical process. The SHG signal depends on the elements of the second-order susceptibility tensor $\chi^{(2)}$ \cite{klingshirn2012semiconductor}, which are non-vanishing for non-centrosymmetric media (i.e. odd number of TMD layers) along the armchair direction of TMDs, see FIG.~\ref{fig1}a. This crystalographic direction can thus be directly determined by rotating the linear polarisation in the experiment. The resulting polar plot of the SHG intensity reveals the crystallographic orientation of the material, useful to precisely measure the relative twist angles (stacking) of homo- and heterobilayers \cite{shinde2018stacking}, see FIG.~\ref{fig3}e. Novel techniques exploit this effect to map with high spatial resolution ($\approx$ 400~nm) the armchair orientation in twisted bilayers \cite{shg2019twist}, as well as in large ($>10^{4}$~$\mu$m$^{2}$) monolayer areas and evaluate their crystal quality since dislocations and grain boundaries can affect the armchair orientation i.e. change the lattice vector \cite{shg2018ultrahigh}. Also the presence of uniaxial strain can be quantified by measuring the SHG intensity along different polarisation directions \cite{mennel2019second}. \\
\indent The SHG response of a material does not only reflect the crystal structure but also depends on the electronic excitations \cite{farenbruch2020magneto}. The efficiency of the SHG signal can be enhanced by several orders of magnitude by selecting the excitation energy to be in resonance with excitonic states of the investigated materials \cite{wang2015giant}. This opens the way for investigating in general the role of electronic excitations (exciton resonances) on the SHG response of a material \cite{seyler2015electrical,PhysRevLett.103.057203}. Using the sensitive SHG response to detect the energy position of electronic transitions is termed 'SHG spectroscopy'. In bilayer CrI$_3$ SHG has been shown to originate from the layered antiferromagnetic order, which breaks both the spatial inversion symmetry and the time-reversal symmetry \cite{sun2019giant} of this centro-symmetric crystal. This makes SHG a highly sensitive probe also for magnetic ordering in layered materials. 


\indent \textbf{Acknowledgements.---}  We acknowledge funding from ANR 2D-vdW-Spin, ANR VallEx, ANR MagicValley, ITN 4PHOTON Marie Sklodowska Curie Grant Agreement No. 721394 and the Institut Universitaire de France. We thank Hans Tornatzky, Delphine Lagarde, Andrea Balocchi, Nadine Leisgang, Hongkun Park and Misha Glazov for stimulating discussions and You Zhou for providing data from \cite{PhysRevLett.124.027401}.\\



\end{document}